\documentclass[twocolumn,showpacs,prl]{revtex4}
\usepackage{graphicx}
        
\begin{document}
        \title{First Order Phase Transition and Phase Coexistence in a 
               Spin-Glass Model}
        \author{Andrea Crisanti$^\star$ and  Luca Leuzzi$^\dagger$}
	\affiliation{$^\star$ Dipartimento di Fisica, Universit\`a di Roma, 
	``La Sapienza'' and INFM unit\`a di Roma I,
         P. le A. Moro 2, 00186, Rome, Italy.\\
	$^\dagger$Instituut voor Theoretische Fysica and FOM, 
	Universiteit van Amsterdam, Valckenierstr. 65, 1018 XE, 
	Amsterdam, The Netherlands.}

        \date{ printout: \today}
        %\date{ printout: 16.04.2002}

\begin{abstract}
We study the mean-field static solution of the 
Blume-Emery-Griffiths-Capel model with quenched disorder, an Ising-spin 
lattice gas with quenched random magnetic interaction. 
The thermodynamics is worked out in the Full Replica Symmetry Breaking
scheme.
The model exhibits a high temperature/low density paramagnetic phase.
When the temperature is decreased or the density increased, the system  
undergoes a phase transition to a Full Replica Symmetry Breaking 
spin-glass phase. 
The nature of the transition can be either of the  second order
 (like in the Sherrington-Kirkpatrick model) or, at temperature below 
a given  critical value (tricritical point), of the first order in the 
Ehrenfest  sense, with a discontinuous jump of the order parameter and 
a latent heat.  In this last case coexistence of phases occurs.

\end{abstract}


\maketitle 
The spin-glass phase (SG) has played and is still playing a central role
in the understanding of disordered and complex systems.
The analysis of mean-field models revealed different possible 
scenarios for the SG phase and the transition to it.
Most of the work, however, has been concentrated on just two 
of them.
In order of appearance in literature the first scenario is described
by a Full Replica Symmetry Breaking (FRSB) solution 
characterized by a continuous order parameter function \cite{P80},
which continuously grows from zero by crossing the transition. 
The prototype model is the Sherrington-Kirkpatrick (SK) model \cite{SK75},
a fully connected Ising-spin with quenched random magnetic interactions.
The second scenario, initially introduced by Derrida \cite{Derrida},
provides a transition with a jump in the order parameter
to a SG One Step Replica Symmetry Breaking (1RSB). 
No discontinuity appear, however, in the thermodynamic functions.
Actually, at the transition to the 1RSB SG phase 
the Edwards-Anderson order parameter can either grow continuously
from zero or jump discontinuously to a finite value. 
The first case includes Potts-glasses with three of four states
\cite{GroKanSom85}, the spherical $p$-spin spin glass model in strong
magnetic field \cite{CS} 
and some spherical $p$-spin spin glass model with a mixture 
of $p=2$ and $p>3$ interactions \cite{Nie95,CriCuk}.
The latter case includes, instead, Potts-glasses with more than four 
states \cite{GroKanSom85},
quadrupolar glass models \cite{GroKanSom85,GolShe85},
$p$-spin interaction spin-glass models with $p>2$ 
\cite{Derrida,Gar85,KWPRA87} and the
spherical $p$-spin spin glass model in weak
magnetic field \cite{CS}.
The models of this class, often referred to as 
``discontinuous spin glasses'' \cite{BCKM}, have been widely investigated
in the last years
because of their relevance for the structural glass transition 
observed in fragile glasses \cite{KWPRA87,CHS}.

In all cases discussed so far the transition is {\it always} continuous
in the Ehrenfest sense. To our knowledge,  the first case of a spin glass
undergoing a genuine first order thermodynamic transition is the so called
Ghatak-Sherrington model (GS) \cite{GSJPC77}.
However, besides the full analysis of the
RS solution, the study of 
the FRSB solution for this model was done only 
close to the continuous transition (SK-like) down to and including the
tricritical point because of its complexity \cite{GSJPC77,LAJPC82}.
We also recall that an exactly solvable model, generalization of 
the Derrida's REM \cite{Derrida}, displaying a first order
phase transition to a SG phase with latent head, was introduced 
by Mottishaw \cite{Mot86}. However, at difference with the 
GS model, the SG phase is 1RSB.

Recently a generalization of the GS model \cite{GSJPC77} has been considered 
in connection with the structural glass transition due to the 
conjectured existence \cite{SNAJPF97} 
of a ``discontinuous'', in the above mentioned sense, 
transition to a 1RSB SG phase. This possibility has risen
 new interest in such model
and its finite dimensions version has been
numerically investigated in a search for evidence of
a structural glass transition scenario \cite{dCC}.

To clarify this issue and its compatibility with previous
results on the GS model, we have investigated the whole 
phase diagram, deep in the SG phase, for the  
mean field quenched disorder 
variant of the Blume-Emery-Griffiths-Capel 
model \cite{BEG71} (BEGC), introduced for the $\lambda$ transition 
in mixtures  of He$^3$-He$^4$, which includes the GS model.

In literature there exists two different versions of the model. The
first one is the direct generalization of the original BEGC model
and uses spin-$1$ variables $\sigma_i=-1,0,1$ on each site $i$ of a lattice
\cite{dCNYJPA97,SEPJB99}, while the 
second formulation is a lattice gas ($n_i=0,1$) of spin-$1/2$ variables
($S_i=-1,1$) \cite{ANSJPF96,SNAJPF97}. In both cases the spin variables
interact through quenched random couplings. 
The two formulations are equivalent, at least as far as static properties are
concerned. Indeed, 
by imposing  $\sigma_i\equiv S_i\,n_i$ the 
two models can be transformed one into the other, 
apart from a rescaling of the chemical potential/crystal field
\cite{footnote1}.
In this paper we will use the second formulation
described by the Hamiltonian (DBEGC) \cite{SNAJPF97}
\begin{equation}
\mathcal{H}= - \sum_{i<j} J_{ij} S_i S_j n_i n_j 
  -\frac{K}{N} \sum_{i<j} n_i n_j
    -\mu \sum_i n_i 
%- h \sum_i S_i n_i ,
\label{eq:H}
\end{equation}
representing an Ising spin glass lattice gas 
coupled to a spin reservoir.
The symmetric couplings $J_{ij}$ are quenched Gaussian random variables  
of zero mean 
%${\overline{J_{ij}}}=J_0/N$
 and variance
${\overline{J_{ij}^2}}={\overline{J_{ji}^2}}=J^2/N$. Here and in the 
following  the overline denotes average with respect to disorder.
Limiting cases of the model are the SK model \cite{SK75} obtained for 
$\mu/J\to \infty$,  the site frustrated percolation model \cite{CJPIV93} 
for $K=-1$ and $J/\mu\to \infty$, and the GS model for $K=0$.

To keep the level of the presentation as general as possible, we shall 
avoid technical details and report only the main results for the phase 
diagrams for the three relevant cases of the GS model ($K=0$), 
the frustrated Ising lattice gas ($K=-1$)
and the case of attracting particle-particle interaction ($K=1$).

Applying the standard replica method, the FRSB solution in the SG phase
is described by the order parameter function \cite{P80}
\begin{equation}
   q(x)=\int_{-\infty}^{\infty} {\rm d}y\, P(x,y)\, m(x,y)^2 \ ,
\label{eq:q}
\end{equation}
and the density of occupied sites by
\begin{equation}
  \rho= \int_{-\infty}^{\infty} dy\ P(1,y)\ 
           \frac{\cosh \beta J y}{e^{-\Theta_1}+\cosh \beta J y} \ ,
\label{eq:d}
\end{equation}
where 
$\Theta_1\equiv (\beta J)^2\, [\rho-q(1)] / 2 +\beta(\mu+K \rho)$
and
$m(x,y)$ and $P(x,y)$~\cite{SDJPC84} are solution of
\begin{equation}
\dot m(x,y)= -\frac{\dot q}{2} m''(x,y)+\dot\Delta(x)\, m(x,y)\, m'(x,y) \ ,
\label{eq:meq}
\end{equation}
\begin{equation}
  \dot P(x,y)=\frac{\dot q(x)}{2}P''(x,y)
           +\dot\Delta(x)\left[P(x,y) m(x,y)\right]' \ , 
\label{eq:Peq}
\end{equation}
with boundary conditions
\begin{eqnarray}
&&m(1,y)= \frac{\sinh(\beta y)}{e^{-\Theta_1}+\cosh(\beta y)}\ ,
\\
&&P(0,y)=\frac{1}{\sqrt{2\pi q(0)}}
\exp\left\{- \frac{y^2}{2 q(0)}\right\} \ .
\end{eqnarray}
The functions $m(x,y)$ and $P(x,y)$ are,
 respectively, the local magnetization and local field
probability distribution at ``time scale'' $x \in[0,1]$ \cite{SDJPC84}, while
$\Delta(x)$ is the Sompolinsky's anomaly \cite{SPRL81}. As usual,
the ``dot'' denotes partial derivative with respect to $x$ while
the ``prime'' the one with respect to $y$.

\begin{figure}[hbt!]
\begin{center}
{
 \includegraphics*[height=5.4 cm]{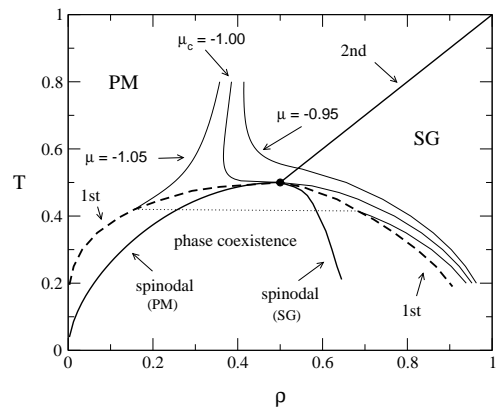}
}
\end{center}
\vskip -0.7 cm
\caption{$T-\rho$ phase diagram of the DBEGC for $K=1$. The dot marks the 
         tricritical point $\mu_c = -1$, $T_c=1/2$, $\rho_c=1/2$. 
	See text for discussion.
        }
\label{fig:phd.K1.0_Tr}
\vskip -0.3 cm
\end{figure}

\begin{figure}[ht!]
\begin{center}
{
\includegraphics*[height=5.4 cm]{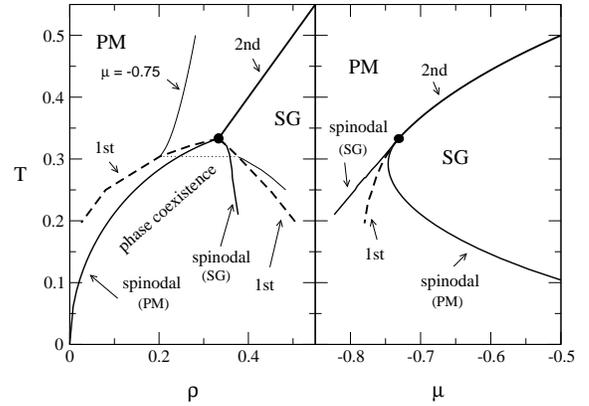}
}
\end{center}
\vskip -0.7 cm
\caption{$T-\rho$ and $T-\mu$ phase diagrams for $K=0$.
        A line at constant $\mu=-0.75<\mu_c$ is shown in the $T-\rho$ plane.}
\label{fig:phd.K0.0_Tr}
\vskip -0.3 cm
\end{figure}

\begin{figure}[t]
\begin{center}
{
\includegraphics*[height=5.4 cm]{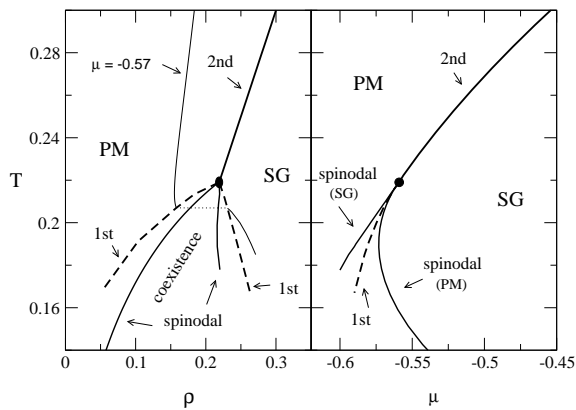}
}
\end{center}
\vskip -0.7 cm
\caption{$T-\rho$ and $T-\mu$ phase diagrams for $K=-1$. In the $T-\rho$ plane
         the line at constant $\mu=-0.57<\mu_c$ is plotted. 
	}
\label{fig:phd.K-1.0_Tr}
\vskip -0.3 cm
\end{figure}

All thermodynamic quantities can be written in terms of the above functions.
For example, defining $\tilde{K}\equiv K+\beta J/2$, 
the internal energy density $u$ and the entropy density $s$ 
read
\begin{equation}
u=- \frac{\tilde{K}}{2} \rho^2
-\mu\  \rho +\frac{\beta J^2}{2} q(1)^2
+\int_0^1 dx~q(x)~\dot\Delta(x) \ ,
\label{eq:u_d_i}
\end{equation}

\begin{eqnarray}
  s &=& -\rho\Theta_1
    -\frac{(\beta J)^2}{4}\left[\rho-q(1)\right]^2 
   + \int_{-\infty}^\infty {\rm d}y\, P(1,y)
\nonumber \\
	&&\times\left\{\log\left[2+2~e^{\Theta_1}\cosh(\beta J y)\right]
           - \beta Jy\,m(1,y)\right\} \ .
\label{eq:entro} 
\end{eqnarray}
We have solved the coupled equations (\ref{eq:q})-(\ref{eq:Peq}) 
in the Parisi's gauge $\dot\Delta=-\beta J x \dot q(x)$
using the pseudo-spectral method introduced in \cite{CLP}.
Analyzing the stability of the RS solution one gets
the critical lines
\begin{eqnarray}
&&1-(\beta J\rho)^2=0 \ ,
\label{eq:L0}
\\
&&1-\beta\tilde{K}(1-\rho)\rho=0 \ ,
\label{eq:L1}
\end{eqnarray}
above which
the only solution is the paramagnetic (PM) solution $q(x)\equiv 0$
for $x\in[0,1]$,
$\rho = 1 / [ 1 + e^{-\Theta_1}]$. This is stable for 
any value of $K$. In the $T\,-\,\rho$ plane, these are, 
respectively, the straight line and the 
left branch of the spinodal line shown, in Figs.
\ref{fig:phd.K1.0_Tr}, \ref{fig:phd.K0.0_Tr} and \ref{fig:phd.K-1.0_Tr}
(for $K=1, 0, -1$, respectively).
The two lines meet at the
tricritical point 
\begin{eqnarray}
&&T_c=\rho_c=\frac{-3/2+K+ \sqrt{K^2-K+9/4}}{2 K} \ ,
\label{eq:trc}
\\
&&  \mu_c= -\frac{1}{2}-\rho_c\left[
	K+\log\left(\frac{1}{\rho_c}-1.\right) 
	\right] \ .
\label{eq:muc}
\end{eqnarray}

By crossing the critical line (\ref{eq:L0}) above the tricritical point
($\rho > \rho_c$, $T>T_c$,  $\mu> \mu_c$ )
the system undergoes a continuous phase transition
of the SK-type to a FRSB SG phase, with a non-trivial continuous 
order parameter function $q(x)$ which smoothly
grows from zero. 
\begin{figure}[ht!]
\begin{center}
{
\includegraphics*[height=5.4 cm]{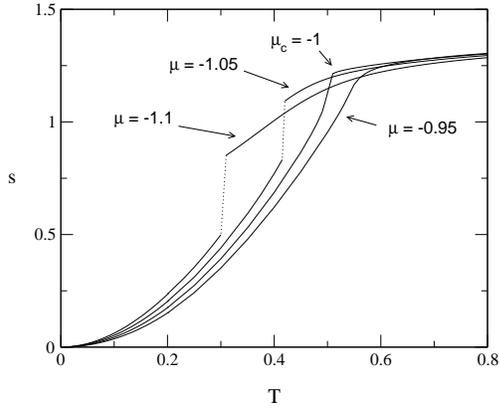}
}
\end{center}
\vskip -0.7 cm
\caption{Entropy density as a function of temperature for $K=1$.
        For $\mu<\mu_c=-1$
	the entropy is discontinuous at the transition temperature.
        }
\label{fig:ent.K1.0}
\vskip -0.3 cm
\end{figure}

\begin{figure}[ht!]
\begin{center}
{
\includegraphics*[height=5.4 cm]{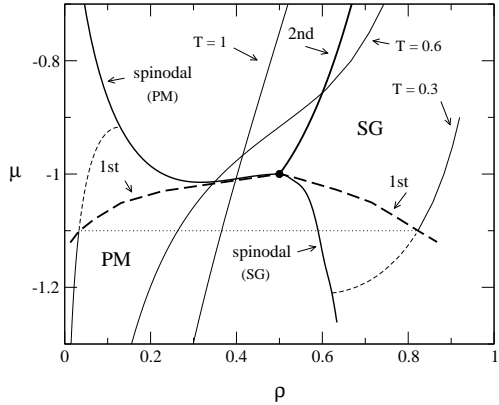}
}
\end{center}
\vskip -0.7 cm
\caption{$\mu-\rho$ phase diagram of the DBEGC for $K=1$. Three 
isothermal lines are plotted, 
two above and one below the tricritical temperature
$T_c=1/2$. For $T=0.3$ also the metastable branches are shown, both in the RS 
PM phase and in the FRSB SG phase. They reach the spinodal lines with zero 
derivative. In this plane of conjugated thermodynamic variables a 
Maxwell construction can be explicitly performed.
        }
\label{fig:phd.K1.0_Mr}
\vskip -0.3 cm
\end{figure}

\begin{figure}[ht!]
\begin{center}
{
\includegraphics*[height=5.4 cm]{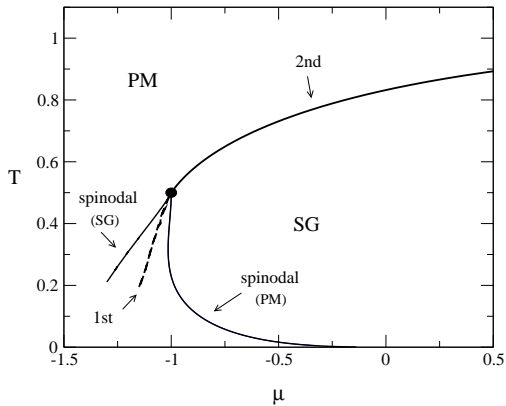}
}
\end{center}
\vskip -0.7 cm
\caption{$T-\mu$ phase diagram of the DBEGC for $K=1$. 
        }
\label{fig:phd.K1.0_TM}
\vskip -0.3 cm
\end{figure}

\begin{figure}[ht!]
\begin{center}
{
\includegraphics*[height=5.4 cm]{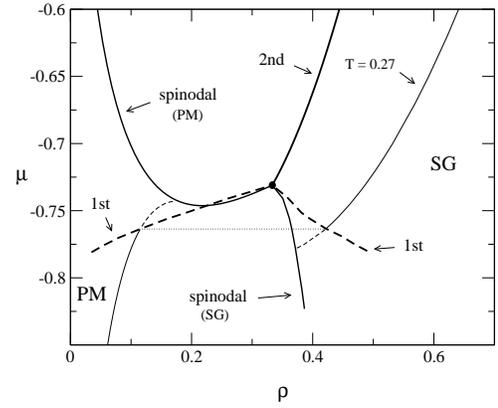}}
\end{center}
\vskip -0.7 cm
\caption{$\mu-\rho$ phase diagram of the DBEGC for $K=0$.  The dot marks the 
         tricritical point $\mu_c = -0.731$, $T_c=1/3$, $\rho_c=1/3$. 
          The isothermal at $T=0.27$ is plotted, together with its metastable 
parts (dotted line) both in the SG and in the PM phase. }
\label{fig:phd.K0.0_Mr}
\vskip -0.3 cm
\end{figure}

\begin{figure}[ht!]
\begin{center}
{
\includegraphics*[height=5.4 cm]{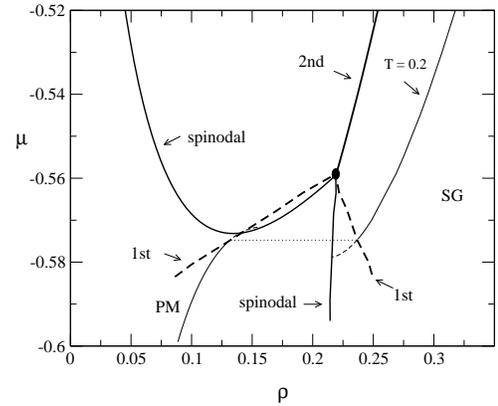}
}
\end{center}
\vskip -0.7 cm
\caption{$\mu-\rho$ phase diagram of the DBEGC for $K=-1$, with isothermal 
	at $T=0.2<T_c$. The dot marks the 
         tricritical point     $\mu_c = -0.559$, $T_c=0.219$, $\rho_c=0.219$. 
        }
\label{fig:phd.K-1.0_Mr}
\vskip -0.3 cm
\end{figure}
Below the tricritical point the scenario is completely different
with a  transition from the PM phase to
a FRSB SG phase with $q(x)$ which discontinuously 
jumps from zero to a non-trivial (continuous) function.
 At the critical temperature the 
 entropy is discontinuous, see Fig. \ref{fig:ent.K1.0}, 
and hence a latent heat
is involved in the transformation,
implying that the transition is of the 
 first order in the Ehrenfest sense.
The transition line is determined by the free energy balance between
the PM and the SG phase \cite{Mot86}, and is shown as a
broken line in the phase diagrams.
The line (\ref{eq:L1}) where the
PM solution becomes unstable, and the equivalent line 
from the SG side are the {\it spinodal} lines. This can be 
better appreciated in the $\mu\,-\,\rho$ plane.
From Fig. \ref{fig:phd.K1.0_Mr} we indeed see that the isothermal lines
cross the instability lines with zero derivative and hence a diverging
compressibility 
$\kappa =  (1/ \rho^2)  \partial \rho/\partial \mu.$
It can be shown that the first order transition line can be 
determined in the $\mu\,-\,\rho$ phase diagram from the isothermal and 
spinodal lines by using a Maxwell construction.
In the region between the first order transition line and the spinodal line the
pure phase is metastable.
Below the spinodal lines (in the $T\,-\,\rho$ plane) no pure phase can exist
and the system is in a mixture of PM and SG phase 
({\it phase coexistence}).
Finally, the phase diagram in the $T\,-\,\mu$ plane, for $K=1$, is shown
in Fig. \ref{fig:phd.K1.0_TM}.

By varying $K$ the scenario remains qualitatively unchanged. 
The only effect of a strong repulsive particle-particle interaction is to 
increase the phase diagram zone where the empty system ($\rho=0$) is the only 
stable solution.
In order to find further phases, e.g. an antiquadrupolar phase \cite{SNAJPF97},
a generalization of the present analysis to a two component 
magnetic model \cite{KSSPJETP85}, including  quenched disorder, has to be
carried out \cite{CL2}.
In Figs. \ref{fig:phd.K0.0_Tr}-\ref{fig:phd.K-1.0_Mr} 
we show the phase diagrams
for $K=0$, the GS model \cite{GSJPC77}, and
$K=-1$ the frustrated lattice gas \cite{dCC}.

In conclusion, we have discussed the complete phase diagram 
of the DBEGC model in the mean field limit, explicitly solving
the FRSB equations in the whole SG phase
with the pseudo-spectral method developed in 
Ref. \cite{CLP}.
Our results rule out the possibility of a 1RSB phase: the SG phase is 
{\em always} of FRSB type.
The transition 
between the PM phase and the SG phase can be either of the SK-type
or, below the tricritical temperature, a first order
thermodynamic phase transition.
In the latter case,  like in the gas-liquid 
transition, a  latent heat is involved in the transformation.
Moreover, for a certain range of parameters (between the spinodal lines), 
no pure  phase is achievable, not even as a metastable one, and the two phases 
coexist.

\noindent
A.C. acknowledges support from the INFM-SMC center.


\begin{thebibliography}{99}
\bibitem{footnote1}
The transformations is $\sigma_i=S_i~n_i$ and 
$D=-\mu+T\log 2$, where $D$ is the 
crystal field.
\end{thebibliography}
\end{document}